\documentclass[onecolumn, 12pt, draftclsnofoot, comsoc]{IEEEtran}

\usepackage{cite}
\usepackage{amsmath,amssymb,amsfonts}
\usepackage{graphicx}
\usepackage{xcolor}
\usepackage{kotex}  

\usepackage{caption}
\usepackage{subcaption}

\begin{document}
%
\title{\Huge Rate-Splitting\hspace{-1mm} Multiple\hspace{-1mm} Access\hspace{-1mm} for\hspace{-1mm} 6G\hspace{-1mm} Networks: Ten Promising Scenarios and Applications}
%
%
%

\author{ Jeonghun~Park, 
       Byungju Lee,  Jinseok Choi, Hoon~Lee, Namyoon~Lee, 
         Seok-Hwan~Park, Kyoung-Jae Lee,  Junil~Choi, 
        Sung~Ho~Chae, Sang-Woon Jeon, Kyung~Sup~Kwak, 
       Bruno Clerckx, and  Wonjae~Shin
        \thanks{
J. Park is with Yonsei University, Seoul, S. Korea;
B. Lee is with Incheon National University, Incheon, S. Korea;
J. Choi is with KAIST, Daejeon, S. Korea;
H. Lee is Pukyong National University, Busan, S. Korea; 
N. Lee is Korea University, Seoul,  S. Korea; 
S.-H. Park is Jeonbuk National University, Jeonju, S. Korea;
K.-J. Lee is Hanbat National University, Daejeon, S. Korea;
J. Choi is KAIST, Daejeon, S. Korea;
S. H. Chae is Kwangwoon University, Seoul, S. Korea; 
S.-W. Jeon is Hanyang University, Ansan, S. Korea;
K. S. Kwak is Inha University, Incheon, S. Korea;
B. Clerckx is Imperial College London, U.K, and also with Silicon Austria Labs (SAL), Graz, Austria;
W. Shin is Ajou University, Suwon, S. Korea, and also with Princeton University, NJ, USA.
}}

%
%

\markboth{Submitted to IEEE Network}%
{Shell \MakeLowercase{\textit{et al.}}: Bare Demo of IEEEtran.cls for IEEE Journals}

\maketitle

\vspace{-0.5cm}
\begin{abstract}
In the upcoming 6G era, multiple access (MA) will play an essential role in achieving  high throughput performances required in a wide range of wireless applications. Since MA and interference management are closely related issues, the conventional MA techniques are limited in that they cannot provide near-optimal performance in universal interference regimes. Recently, rate-splitting multiple access (RSMA) has been gaining much attention. {RSMA splits} an individual message into two parts: a common part, decodable by every user, and a private part, decodable only by {the} intended user. Each user  first decodes the common message and then decodes { its} private message by applying successive interference cancellation (SIC). By doing so, RSMA not only embraces the existing MA techniques as special cases but also provides significant performance gains by efficiently mitigating inter-user interference in a broad range of interference regimes. In this article, we first present the theoretical foundation of RSMA. Subsequently, we put forth four key benefits of RSMA: {\it spectral efficiency, robustness, scalability}, and {\it flexibility}. Upon this, we describe how RSMA can enable ten promising scenarios and applications  along with future research directions to pave the way for 6G. 
\end{abstract}


\IEEEpeerreviewmaketitle

\section{Introduction}
Due to the broadcast nature of the wireless medium, a multi-user system is the natural appearance of modern wireless communications. Accordingly, how to serve multiple users { via} multiple access (MA) techniques { plays} a pivotal role in achieving high spectral efficiency, which is {undoubtedly a} key requirement in 6G. 
In multiple-input multiple-output (MIMO) systems, spatial division multiple access (SDMA) that exploits abundant spatial degrees-of-freedom (DoFs) has become a mainstream MIMO-MA. From an information-theoretic perspective, SDMA can be classified as a treating interference as noise (TIN) type of interference management strategy. Since TIN is optimal in a weak interference regime, SDMA provides near-optimal spectral efficiency performance when the inter-user interference is {weak}, e.g., in underloaded systems { with orthogonal users channels and} perfect channel state information at transmitters (CSIT). Unfortunately, practical wireless systems often encounter severe interference. For example, perfect CSIT cannot be { guaranteed} in downlink due to limited feedback or pilot contamination. In such a case, TIN is far from optimal; which encourages the use of more aggressive types of interference management schemes. Motivated by this, as a joint decoding (JD) type of interference management strategy, MIMO non-orthogonal multiple access (NOMA) has been proposed \cite{shin:commmmag:17}. 
Nonetheless, in MIMO, it is infeasible to obtain the optimal SIC order because each user's spectral efficiency is determined as a complicated function of channel and beamforming vectors. This severely degrades the performance of MIMO-NOMA, even worse than SDMA \cite{Bruno:arxiv:22}. 

The lessons { learned from} SDMA and MIMO-NOMA motivate the development of a new MA technique that can embrace both TIN and JD types of interference management strategies. Recently, rate-splitting multiple access (RSMA) has been proposed as a promising MIMO-MA technique \cite{Bruno:arxiv:22}. A key idea of RSMA is splitting each user's message into two parts, each of which is a common and a private part, respectively. Each common part is combined and jointly encoded {into} a common message by using a shared codebook. {Each private part is encoded using a private codebook, known to the corresponding user as in SDMA.} {With} this unorthodox message construction, each user decodes and removes the common message first, then decodes the private message through SIC. 
By adjusting the common message's {rate, power, and content}, RSMA can cover SDMA (TIN) or MIMO-NOMA (JD) as a special case \cite{Bruno:arxiv:22}. This feature endows RSMA with efficiency, robustness, scalability, and flexibility in several environments and also makes RSMA a key enabler for promising 6G applications and scenarios. In this article, we articulate where the RSMA gains come from, which benefits can be obtained, and how these benefits enable various 6G scenarios and applications.

\section{Theoretical Foundation of RSMA: What Drives RSMA} \label{sec:2}



A key feature of RSMA is to split individual users' messages into a common and a private part, thereafter to construct a common message by jointly encoding the common parts. Subsequently, applying SIC, each user decodes the common message first and removes it; then decodes the private message. Provided that the common message rate is properly determined, the spectral efficiency gains are achieved by mitigating the inter-user interference. One may confuse RSMA with NOMA or multicasting. Indeed, RSMA is similar to NOMA in that SIC is used on the user side. Further, RSMA is similar to multicasting in that a common message aimed to be decoded by every user is used. Nonetheless, they are fundamentally different in several aspects, and understanding these differences is key to capturing the theoretical foundation of RSMA. 

In NOMA, a user fully decodes other users' messages, which is classified into a JD type of strategy. In this type of strategy, it is of importance to use the optimal SIC order because the information rate of a message is determined as the {\it{minimum value}} among the spectral efficiencies of users allowed to decode the corresponding message. 
In SISO, it is straightforward to find the optimal SIC order since the spectral efficiency is only defined in the power domain. In MIMO, however, due to the spatial domain, it is difficult to identify the optimal SIC order. The use of unsuitable SIC order causes severe performance degradation even below SDMA. Additionally, NOMA has no flexibility in message construction, i.e., it cannot be reduced to conventional SDMA depending on channel conditions. 
{In contrast}, RSMA composes a common message that includes portions of every user's message, by which the SIC order is greatly simplified into two steps. 
This feature also makes RSMA resilient. For instance, if the common message rate is zero, RSMA boils down to SDMA. If the common message is composed only of a particular user's common part and the corresponding user does not have a private message, then RSMA becomes NOMA. In this sense, RSMA embraces both TIN and JD types of strategies, and thus RSMA is more versatile to adapt various channel conditions. 

\begin{figure*}
\centering 
\includegraphics[width=6.5in]{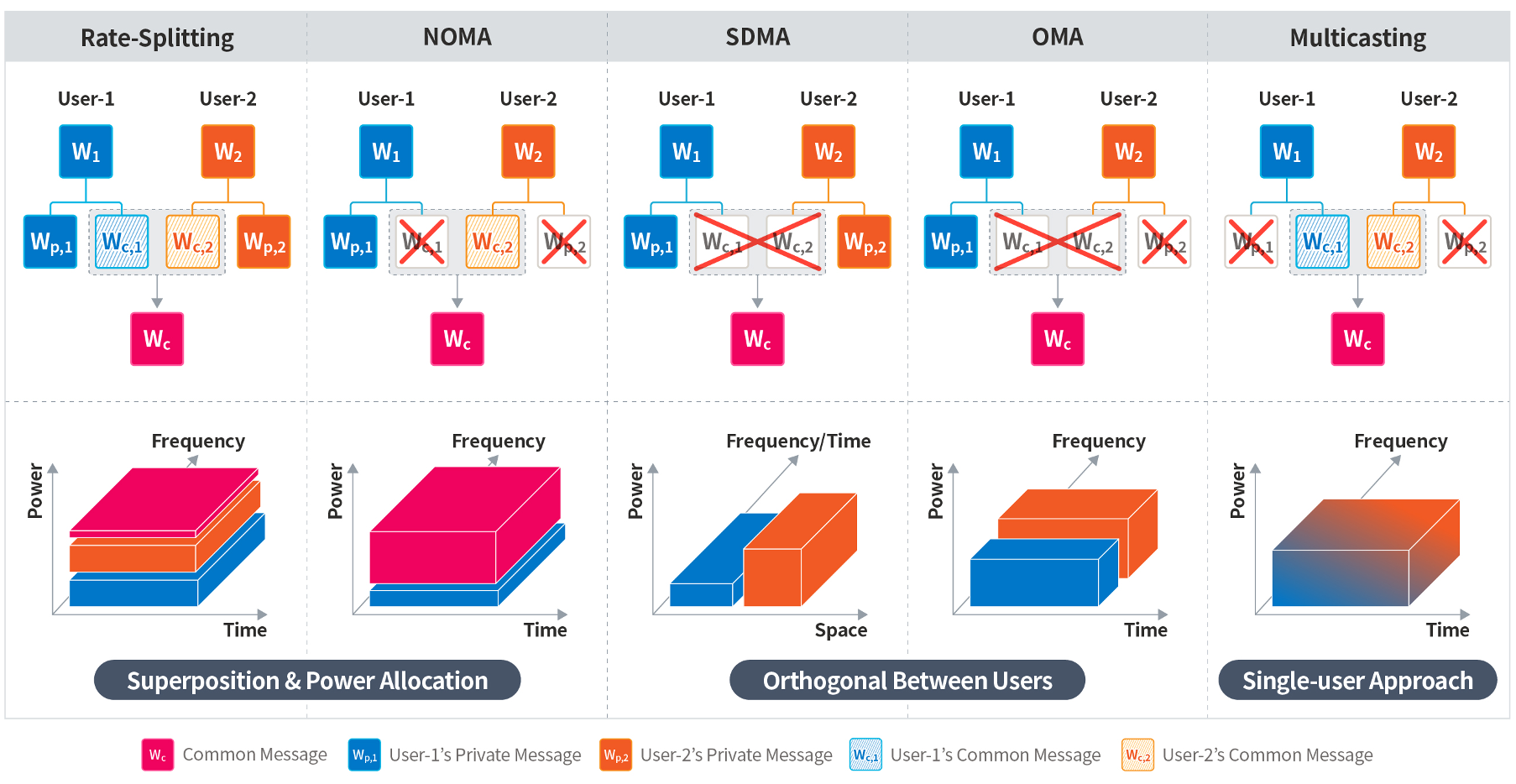} \vspace{-0.7cm}
\caption{{Message construction and resource allocation of various MA techniques from the viewpoint of RSMA.}} \label{fig:concept2} \vspace{-0.3cm}
\label{fig:message_construction}
\end{figure*}


RSMA is also different from multicasting, where a key difference is in the construction of a common message. In multicasting, a common message is made by the service demand. That is to say, a common message in multicasting is actually required for every user in a network, e.g., live video streaming or news feed. {In contrast to that,} in RSMA, a common message is artificially constructed with the aim {of mitigating interference}. For this reason, RSMA is also suitable in a typical unicast message scenario, wherein each user needs different individual messages. 
To help understand the difference between RSMA and other MA techniques, we provide an illustration of {message construction along with resource allocation strategies} in Fig~\ref{fig:message_construction} {for} a two-user case.

Now, we rationalize the performance gains of RSMA from an information-theoretic viewpoint. Consider a two-user vector broadcast channel in which a transmitter equipped with multiple antennas serves two single-antenna users. In this channel, when perfect CSIT is available, sending two independent messages with linear precoding (i.e., SDMA) is optimal within a constant gap to the capacity. 
Nevertheless, in the presence of CSIT error, inter-user interference is inevitable. This naturally leads us to interpret the two-user vector broadcast channel through a lens of a classical two-user interference channel. 
In particular, after applying linear beamforming with imperfect CSIT, an equivalent channel model is a two-user interference channel with limited transmitter cooperation by means of sharing two messages without CSIT sharing (since we do not know CSIT error). Consequently, the classical two-user interference channel can be considered a degraded version of the interference channel with limited cooperation. This interpretation allows greatly facilitating the analysis and the gleaning of insights on RSMA.

Unfortunately, the capacity of this simplest two-user interference channel with partial transmit cooperation is still unknown. Notwithstanding the lack of a capacity-achieving scheme, RSMA allows for attaining spectral efficiency gains with the Han-Kobayashi scheme known as quasi-optimal in the classical two-user interference channel \cite{etkin:tit:08}. This interpretation justifies gains of RSMA in MIMO broadcast channels with imperfect CSIT over the existing SDMA or NOMA. 

In line with our interpretation, we note that SDMA and NOMA provide near-optimal performance at some particular points. For instance, SDMA is quasi-optimal when the inter-interference amount is very weak, or NOMA is quasi-optimal when the channel of the one interfering link is very strong. As mentioned, by adjusting the common message's {rate, power, and content}, RSMA embraces SDMA (allocating zero power to the common message) or NOMA (composing the common message with a single user's message), which brings significant gains to RSMA in universal interference regimes \cite{Bruno:arxiv:22}. 




\section{Ten Promising Scenarios and Applications for 6G} \label{sec:3}


{{In this section, we put forth ten representative scenarios for 6G networks and explain how RSMA is applied to enable these scenarios. 
Depending on the system environment, RSMA gains are manifested into four particular benefits: spectral efficiency, robustness, scalability, and flexibility. 
For better clarity and understanding, we categorize the ten scenarios into four distinct classes depending on the RSMA benefits. Upon this categorization, we account for the main technical challenges given to each scenario and how RSMA addresses these issues. }}



\begin{figure*}
\centering \vspace{-0.5cm}
\includegraphics[width=7in]{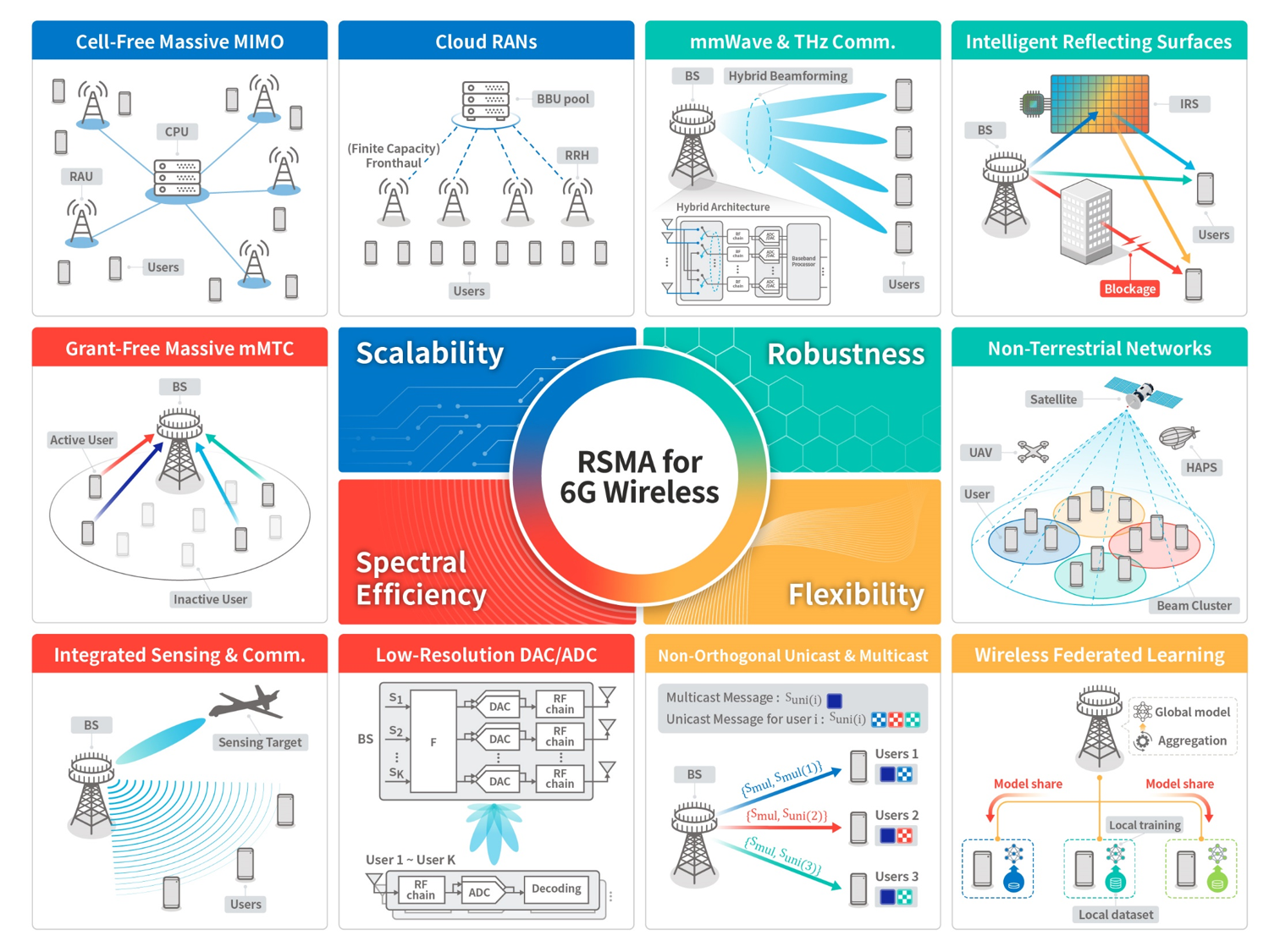} \vspace{-0.1cm}
\caption{Four distinct benefits of RSMA and ten promising 6G applications and scenarios that each benefit is especially useful. } \label{fig:rsma_scenario} \vspace{-0.3cm}
\end{figure*}

{{
{\bf{Spectral Efficiency}}: Some 6G scenarios face {{a lack of available spatial resources}}. For instance, in grant-free massive machine-type communication (mMTC), an excessive number of users, e.g., tens of millions of machine-type devices, attempt to connect to a base station (BS) whose number of antennas is less than the number of devices. This also happens when the spatial resources are occupied for multiple goals such as communication and sensing, i.e., integrated sensing and communication (ISAC), or when the spatial resources are not fully available due to hardware impairments, i.e., low-resolution quantizers. The scarcity of spatial resources incurs a large amount of interference, which cannot be  removed even with perfect CSIT.  
This is where RSMA comes in handy. Applying RSMA, the interference is efficiently reduced by using SIC  thanks to its ability of partially decoding interference and partially treating interference as noise; therefore, achieving high spectral efficiency is still feasible even if spatial resources are not enough. 

{\bf{Robustness}}: Acquiring CSIT is especially difficult in certain 6G scenarios. Specifically, in hybrid architectures with mmWave and Terahertz, CSIT estimation is hard due to the RF hardware limitation. This
is also true in intelligent reflecting surfaces (IRS) where passive devices are deployed, in non-terrestrial
networks (NTN) due to the fast movement of non-terrestrial flying objects and long propagation delays, and in conventional terrestrial systems in the presence of mobile devices and latency in CSI acquisition.
CSIT imperfection induces inter-user interference, which fundamentally limits spectral efficiency performances. With RSMA, the inter-user interference caused by inaccurate CSIT is reduced using SIC, by which it offers robustness to cope with imperfect CSIT through the presence
of a common message. 
Consequently, RSMA is beneficial in mmWave and THz communications, IRS, NTN, and mobile scenarios. 


{\bf{Scalability}}: 
RSMA does not incur excessive system overheads associated with channel estimation and computational burdens for large-scale networks. 
In this sense, RSMA is a favorable MA technique to scale up the network size. Moreover, in contrast to NOMA, the number of SIC layers does not always increase with the number of users. This feature provides scalability. The scalability of RSMA is extremely useful in 6G network scenarios that aim to deploy large-scale networks for enhancing spectral efficiency and coverage. To be specific, cell-free massive MIMO and cloud RAN are particular beneficiaries of RSMA. 

{\bf{Flexibility}}: 
Flexibility is inherently given to RSMA thanks to its message construction process. Since (at least) two different types of messages (common and private) are encoded in RSMA, an encoding framework of RSMA can be applied to heterogeneous message setups. For instance, in wireless federated learning scenarios, it is required to transmit different types of messages aimed to be decoded at different targets, e.g., edge node or cloud server. Similar to this, in non-orthogonal unicast/multicast transmission, unicast and multicast messages are jointly transmitted, where the flexibility of RSMA is helpful.



In Fig.~\ref{fig:rsma_scenario}, we illustrate four benefits that RSMA can provide and which 6G applications and scenarios are enabled with these benefits. Henceforth, we provide more detailed explanations of ten promising 6G applications and scenarios and how RSMA can enable them.
}}


\subsection{Grant-Free Massive Machine-Type Communications}

In the coming era of 6G, mMTC plays a vital role in providing wireless connectivity to tens of billions of machine-type devices, which in general imposes low-complexity, low-power, sporadic transmissions, and various quality of service (QoS) requirements. In the mMTC scenarios, sporadic transmissions are mostly small-size packets emitted from a huge number of uncoordinated nodes, so that grant-free multiple access schemes are needed to reduce signaling overhead, realize simultaneous reliable connectivity, and avoid severe user collisions.


To share the same time-frequency resource blocks with innumerable internet-of-things (IoT) nodes and mMTC devices, the system operates in an overloaded regime, wherein significant inter-user interference is induced. To handle this without involving excessive signaling overheads  as that being performed in grant-based transmissions, non-orthogonal transmission schemes such as NOMA and RSMA have been proposed to realize grant-free multiple access. 
Thanks to the capability to improve spectral efficiency in an overloaded system with low implementation complexity, uplink RSMA has received significant attention. Compared to uplink NOMA with SIC, uplink RSMA with SIC can achieve the full capacity region of {uplink without time-sharing}, which makes uplink RSMA an essential grant-free multiple access strategy to support mMTC \cite{Bruno:arxiv:22}. 
{{One promising future research direction is to study mMTC with RSMA in a finite blocklength regime where perfect SIC cannot be guaranteed.}}


\subsection{Integrated Sensing and Communications}

Integrated sensing and communications (ISAC) has emerged as a new paradigm for future 6G wireless networks
to tackle the increasing need for high-quality wireless connectivity
as well as accurate and robust sensing capability.
ISAC combines both functionalities of wireless communications and remote sensing
into a single system. By doing so, ISAC can make better use of the spectrum, the hardware,
and of the signal processing.
The challenge is that the spectrum is not only to be used efficiently to serve multiple communication users but also to perform the radar sensing functionality. This brings higher requirements in the spectrum and resource management and a need to manage better the interference between communication users and between radar and communication functionalities.

\begin{figure}
	\centering
\includegraphics[width=1.0\textwidth]{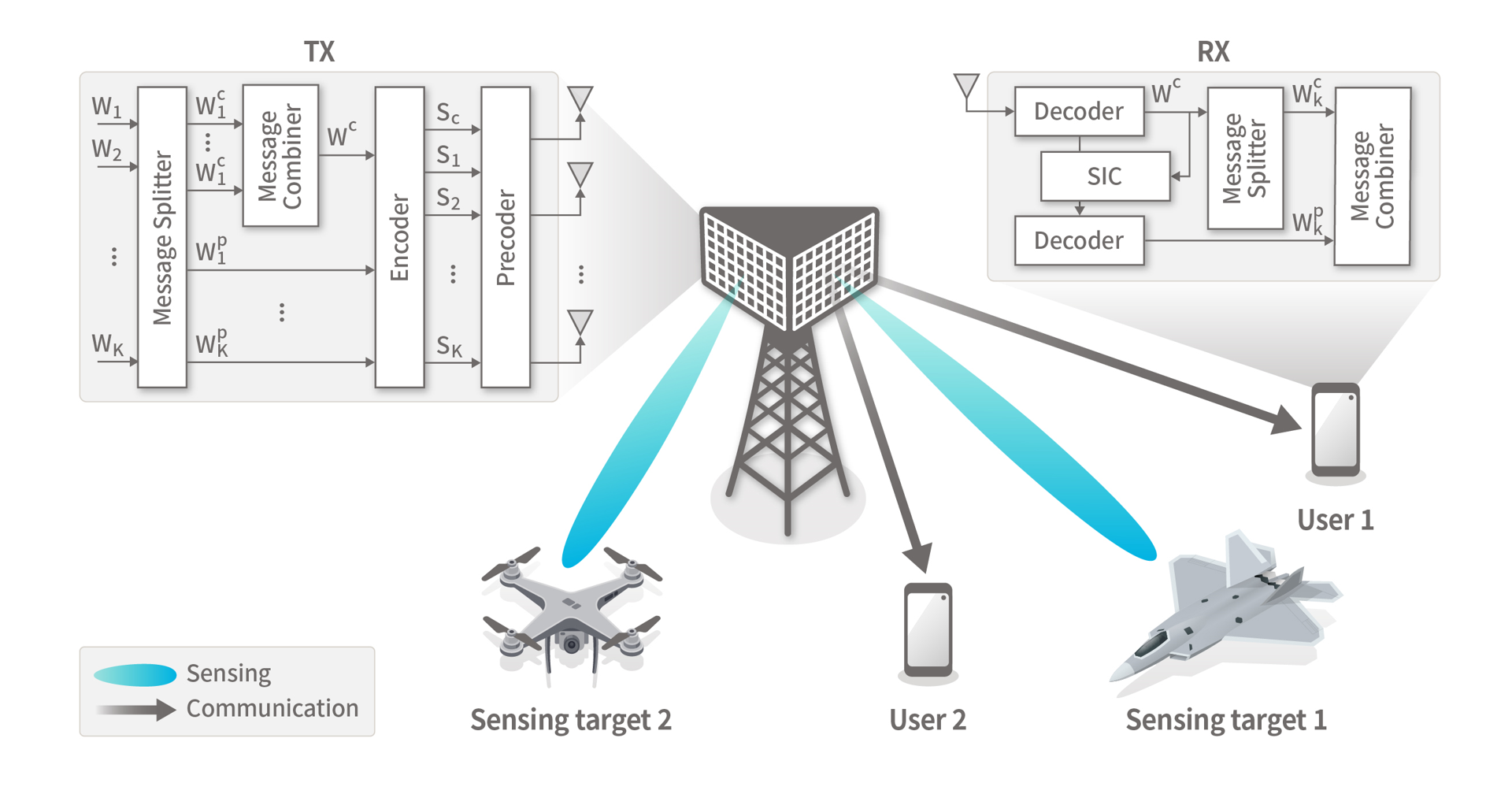}%
	\caption{Model of a RSMA-assisted ISAC system \cite{yin:commlett:22}.}
	\label{RSMA_ISAC}
\end{figure}

RSMA addresses this new challenge thanks to its efficient interference management \cite{yin:commlett:22}. Indeed the common stream $s_c$ in RSMA can be used for the joint purpose of managing inter-user interference (as in RSMA for communication-only network) but also to better sense the environment and conduct radar sensing, as illustrated in Fig. \ref{RSMA_ISAC}. 
This double benefit translates into an enlarged communication-radar tradeoff, characterized for instance in terms of communication rate, e.g. weighted sum-rate (WSR) or max-min fair (MMF) rate, vs. radar metric, e.g. Cramer-Rao Bound (CRB) of the target estimation, beampattern approximation, or radar mutual information. 
In \cite{yin:commlett:22}, it was shown that RSMA significantly outperforms SDMA and NOMA in terms of a tradeoff between communication rate vs. radar metric thanks to its interference management capability. 
{{As future work, incorporating vehicle-to-everything (V2X) or NTN networks into RSMA-assisted ISAC has great potentials.}}


\subsection{RF Impairments and Finite-Resolution Quantizers  } 
Utilizing low-power hardware, such as a low-resolution analog-to-digital converter (ADC) and digital-to-analog converter (DAC), has been regarded as a potential solution to the low-power requirement for 6G IoT applications.
In such low-resolution quantization systems, a key challenge is quantization distortion, whose amount is proportional to the input signal power.



In \cite{park:arxiv:22_2}, it has been shown that RSMA can provide benefits in both increasing spectral efficiency and designing transceiver architectures with low-power hardware. 
When the BS is equipped with low-resolution quantizers, the quantization errors from the common stream as well as the private streams cannot be canceled at the users; resulting in maintaining a similar amount of quantization errors for both RSMA and SDMA. In this regard, the main benefit of using RSMA in low-resolution quantizers comes differently. When considering heterogeneous-resolution quantization systems, it is beneficial to turn off the antennas with low-resolution quantizers in the high SNR regime to suppress the quantization distortion. This phenomenon may lead to  overloaded systems due to the insufficient number of active antennas, and thus RSMA can play a key role in mitigating inter-user interference by leveraging the common stream. 



\begin{figure*}
\centering 
\includegraphics[width=6.8in]{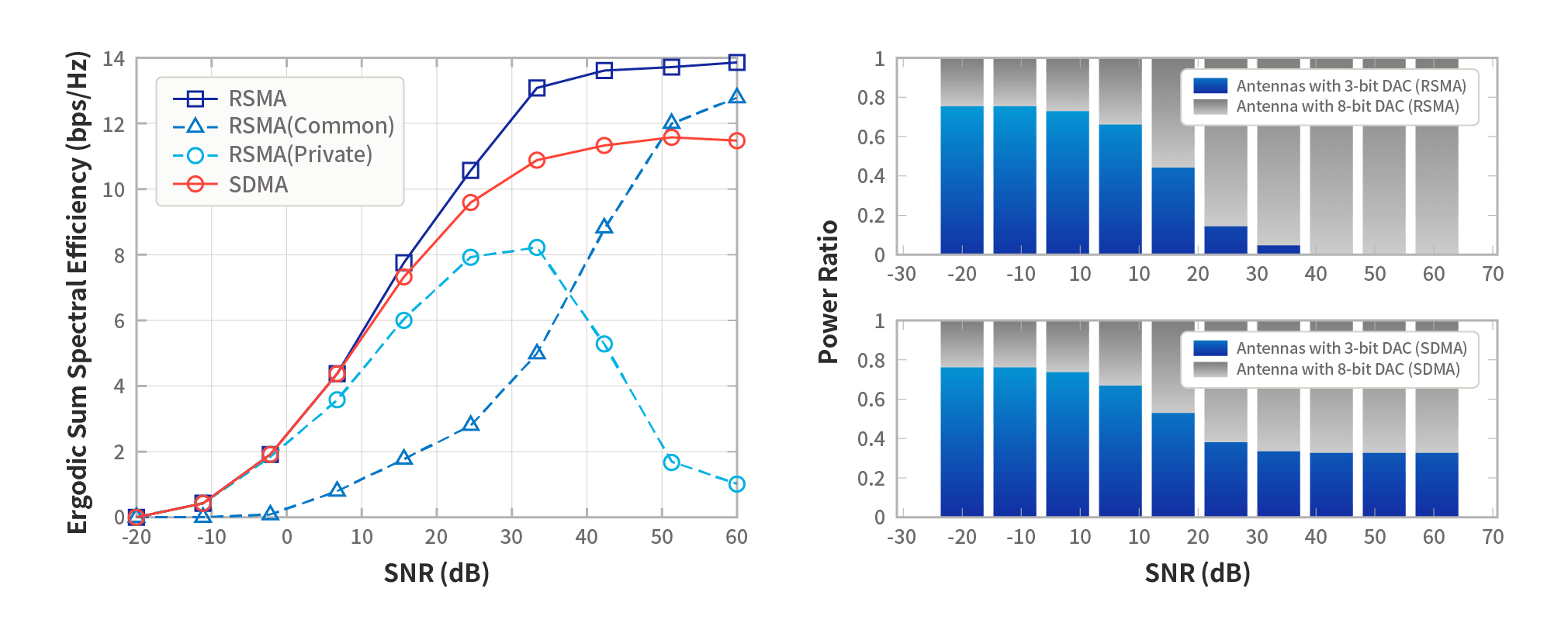} \vspace{-0.1cm}\caption{Sum spectral efficiency and antenna power ratio of RSMA and SDMA for 4 antennas (one with 8-bit DAC and three with 3-bit DAC) and 2 users.} \label{fig:DAC} \vspace{-0.3cm}
\end{figure*}

Fig.~\ref{fig:DAC} verifies such a phenomenon.
It is observed that RSMA tends to use only the antennas with high-resolution quantizers as the SNR increases to avoid large quantization distortion. This leads to the reduction of the spatial DoFs. 
To cope with such insufficient spatial DoFs, RSMA more actively uses the common stream, which helps to handle the inter-user interference effectively. This improves the sum spectral efficiency. 
{{One can further explore the energy efficiency of low-resolution quantizers using RSMA as future work.}}


\subsection{MmWave and THz Communications }

As millimeter-wave (mmWave) has become a key component of 5G communication systems, the THz spectrum may play an important role in the upcoming 6G systems to support extremely high data rate services such as holographic 3D displays or tactile and haptic internet applications.
Although mmWave/THz promises extremely high data rates, it also experiences several challenges, including severe propagation loss and poor penetration and diffraction characteristics, which severely limit the communication distance. To tackle these issues, mmWave \& THz hybrid antenna array structures, in which each subset of antenna elements or sub-array is connected to an antenna port or a radio frequency (RF) chain, allow for the realization of the full potential of multiple beams with much lower cost and power consumption \cite{dai:twc:17}. 

In hybrid array structures, it is difficult to estimate CSIT accurately due to the analog beamforming and combining stage. Specifically, the number of RF chains is typically smaller than the number of analog antennas, so that one can only see the projected channel space. The imperfect CSIT causes inter-user interference, which hinders to improve the spectral efficiency. 
RSMA can be efficiently applied to handle this inter-user interference thanks to the robustness of CSIT estimation error. Exploiting SIC to the common message, RSMA provides considerable spectral efficiency gains for hybrid array architectures in the presence of CSIT estimation error. 
{{As future work, one promising solution is to develop a practical optimization framework for RSMA applicable in hybrid array structures with constant modulus constraints.}}
{{
Besides the hybrid array structure, RSMA also can be a remedy for a coverage issue of mmWave/THz communications. For example, in \cite{cho:twc:23}, a cooperative communication strategy leveraging RSMA was proposed to increase the coverage range in THz communications. 
}}





\subsection{Intelligent Reflecting Surface }
An intelligent reflecting surface (IRS) is generally comprised of a number of passive array elements, wherein each element is tuned to modify phases (or amplitudes) of the impinging signals. 
By doing so, IRS can reflect the incident signals in the desired direction, by which the wireless channels between transmitters and receivers are reshaped. This can resolve a fundamental channel impairment problem by adding an extra propagation path or bypassing an obstacle, e.g., IRS can provide qualified wireless links to a user that has no strong LOS propagation path due to blockages. 
IRS is distinguished from conventional relays in that only passive devices are used to reflect the signals. Specifically, since no active RF chain is deployed in IRS, the consumed energy is much lower than a relay. Further, IRS operates in a full-duplex fashion without inducing self-interference. This makes IRS easy to implement and operate in any wireless environment. Thanks to these, IRS is considered essential, compatible with 6G THz and massive MIMO technologies. 

One possible issue of the IRS is channel estimation \cite{Li:cl22}. Since links between a transmitter-IRS and IRS-receiver are concatenated, perfect knowledge of the CSI of each link is hardly given. This CSI estimation error induces inter-user interference. RSMA is effective in mitigating this interference by applying SIC to a common message for each user. 
In addition to robustness with imperfect CSI, IRS can boost up the common message power. In RSMA, the common message aims to reach every user in a network, therefore a beam shape corresponding to the common message is blunt. This reduces the received signal power of the common message. If multiple IRSs are deployed for each user, each IRS can be designed to increase the common message power received by each user. By doing this, the spectral efficiency of RSMA is further improved. 
{{An open problem may include studying how the RSMA-aided RIS deployments affect a network along with cell-free massive MIMO or cloud-RAN. 
}}

\subsection{Non-Terrestrial Networks }

Non-terrestrial networks (NTN) have emerged as a key enabling technology for  5G  networks and beyond due to their advantages of coverage extension and flexibility through 3D mobility. To achieve universal and seamless connectivity even in rural and remote areas, NTN should not only act as a complement to terrestrial networks (TN) but also cooperatively serve users with TN to achieve virtually no coverage holes. 

\begin{figure*}
\centering 
\includegraphics[width=6.2in]{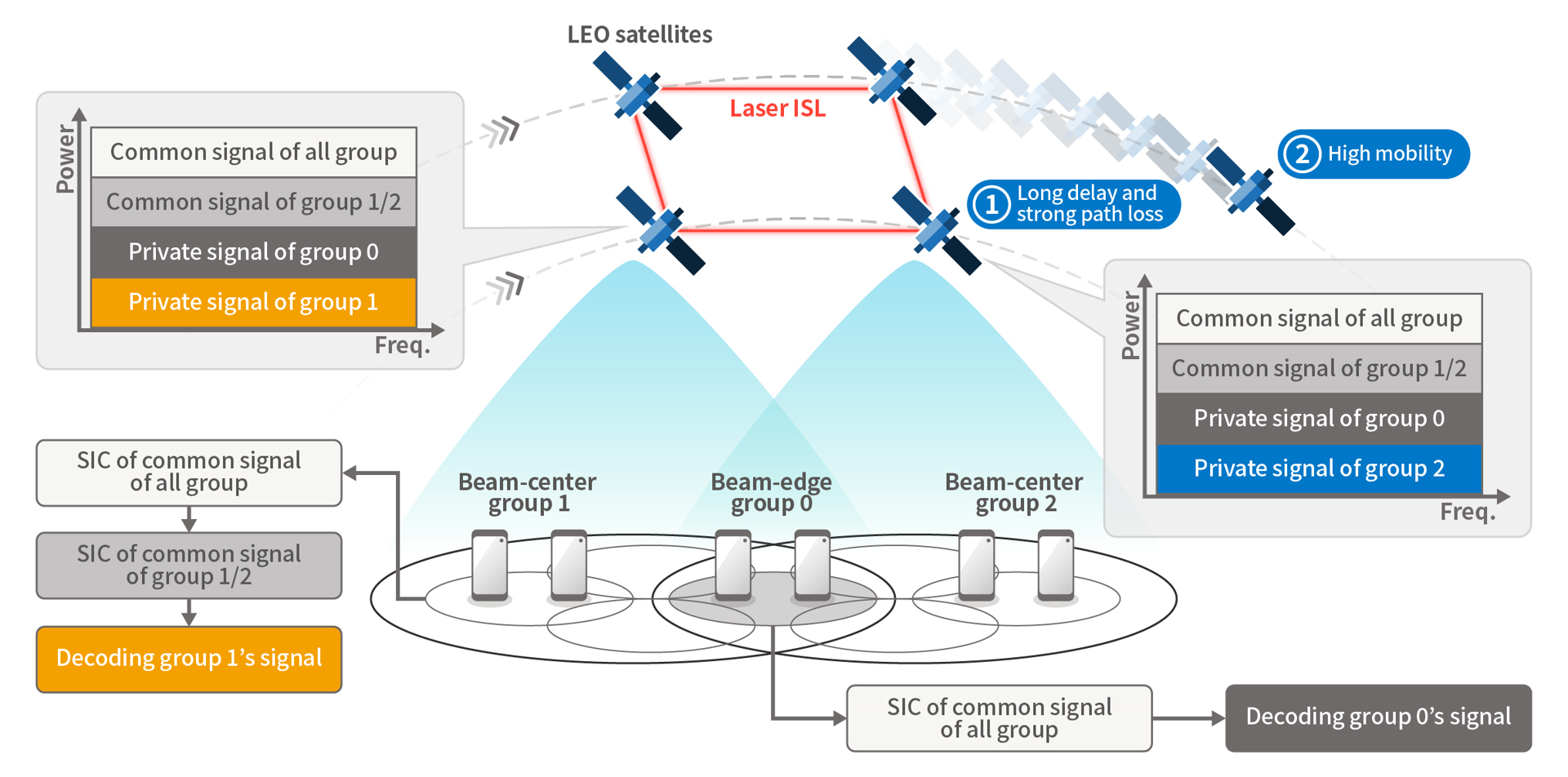} \vspace{-0.1cm}\caption{Rate-splitting-based non-terrestrial networks.} \label{fig:NTN} \vspace{-0.3cm}
\end{figure*}

RSMA can be an effective means of satisfying various users' QoS with different levels of CSI knowledge. In the NTN scenario, perfect CSI is very difficult to obtain with sudden changes in channel environments owing to the relative movement of NTN devices with respect to the ground stations. As illustrated in Fig. \ref{fig:NTN}, RSMA can be applied to multiple beam footprint groups by means of inter-satellite links (ISL). The robustness under imperfect CSI is one of the main advantages of RSMA systems. For example, RSMA was shown to have superiority over SDMA in terms of max-min fair rate in multi-beam satellite systems with imperfect CSI \cite{yin:tcom:21}. 
To further improve the efficacy of the RSMA in the NTN scenario, the following disruptive technologies can be combined. First, IRS can help to resolve the link failure experience in dense urban areas since beams with higher frequencies such as mmWave and THz are vulnerable to blockage. Second, computer vision techniques along with multi-modal sensory data, such as RGB information collected by the camera and LiDAR information, can be applied to identify blockage locations, mobility patterns of NTN terminals, and LOS and non-line-of-sight (NLOS) paths. 
{{One promising research direction includes investigating the high mobility of NTN affects the performance and how to make use of  RSMA as a remedy for the mobility issue.}}

\subsection{Cell-free Massive MIMO }
Cell-free massive MIMO (CFmMIMO) consists of a massive number of distributed access points (APs) that are connected to a central processing unit (CPU) via fronthaul links to serve a large number of users jointly with no boundaries. 
By deploying antennas in a distributed manner, CFmMIMO can provide better spectral and energy efficiencies as well as a more uniform user experience, compared to conventional co-located massive MIMO. Ideal CFmMIMO is free from inter-cell or cluster interference and provides higher coverage probability since APs are brought closer to users. Nevertheless, as the numbers of APs and users become extremely large to cover a huge area, CFmMIMO cannot be scalable to deploy in practice due to the high computational complexity and the fronthaul overhead for channel estimation, power allocation, precoding, and decoding. As a remedy, a dynamic AP clustering can be adopted for a scalable CFmMIMO system. In order to achieve the benefits of the scalable CFmMIMO, interference management, user association, pilot assignment, and transceiver methods must be carefully designed.

RSMA can address the scalability issue in CFmMIMO. 
For instance, in massive MIMO, the interference caused by the pilot contamination can be effectively mitigated for various sounding scenarios by employing RSMA. 
A similar idea is extended to RSMA-enabled CFmMIMO for massive machine-type communications assuming the same pilot signal for all users \cite{mishra:cl:22} with emphasis on precoder design and power allocation. In \cite{flores:icc:22}, the cell-free RSMA shows better performance gains compared with the conventional cell-free SDMA and the conventional co-located RSMA in a practical system with channel imperfection. 
{{For further optimization of RSMA-aided cell-free massive MIMO, it is possible to identify a user group that participates in constructing a common message. 
}}

\subsection{Cloud RAN }

Cloud RAN (C-RAN) is an emerging deployment architecture for 5G/6G that 
centralizes baseband processing functionalities to a virtual baseband processing unit (BBU) pool also referred to as a central processor (CP). 
C-RAN is well suited for ultra-dense networks that improve spectral and energy efficiency through spatial reuse.
The interference signals caused by the dense deployment of BSs can be effectively managed by centralized baseband processing at the CP, which is the most representative benefit of C-RAN. One of the key challenges for realizing the benefits of large-scale C-RAN is associated overheads for acquiring CSIT. The CP requires global CSI in the network to manage the interference, which may contain errors coming from, e.g., imperfect measurement, CSI quantization for feedback or fronthauling, and channel aging due to fronthaul delay. 


RSMA offers scalability to C-RAN by managing the interference caused by scaling-up network size. 
The effectiveness of RSMA in interference mitigation is particularly pronounced in C-RAN systems in which a virtual network MIMO system attempts to overcome a large amount of interference signals.
The robustness of RSMA to inaccurate CSI plays a more important role in C-RAN scenarios where various CSI errors can occur. Another performance bottleneck of C-RAN is the capacity limitation of fronthaul links that connect the CP to BSs. With the flexible rate control among split messages in RSMA, the fronthaul capacity limitation can be better alleviated than in a single-layer SDMA scheme \cite{ahmad:tcom:21}.
Moreover, RSMA can provide additional advantages for C-RAN systems operating with wireless fronthaul. Since the wireless fronthaul link from CP to BSs is an instance of BC, RSMA is in place to increase the fronthaul capacity region compared to conventional multiple access schemes.
{{As future work, joint transmission design for RSMA by incorporating fronthaul capacity is promising. 
}}
%



\subsection{Wireless Federated Learning }

\begin{figure*}
\centering 
\includegraphics[width=6.2in]{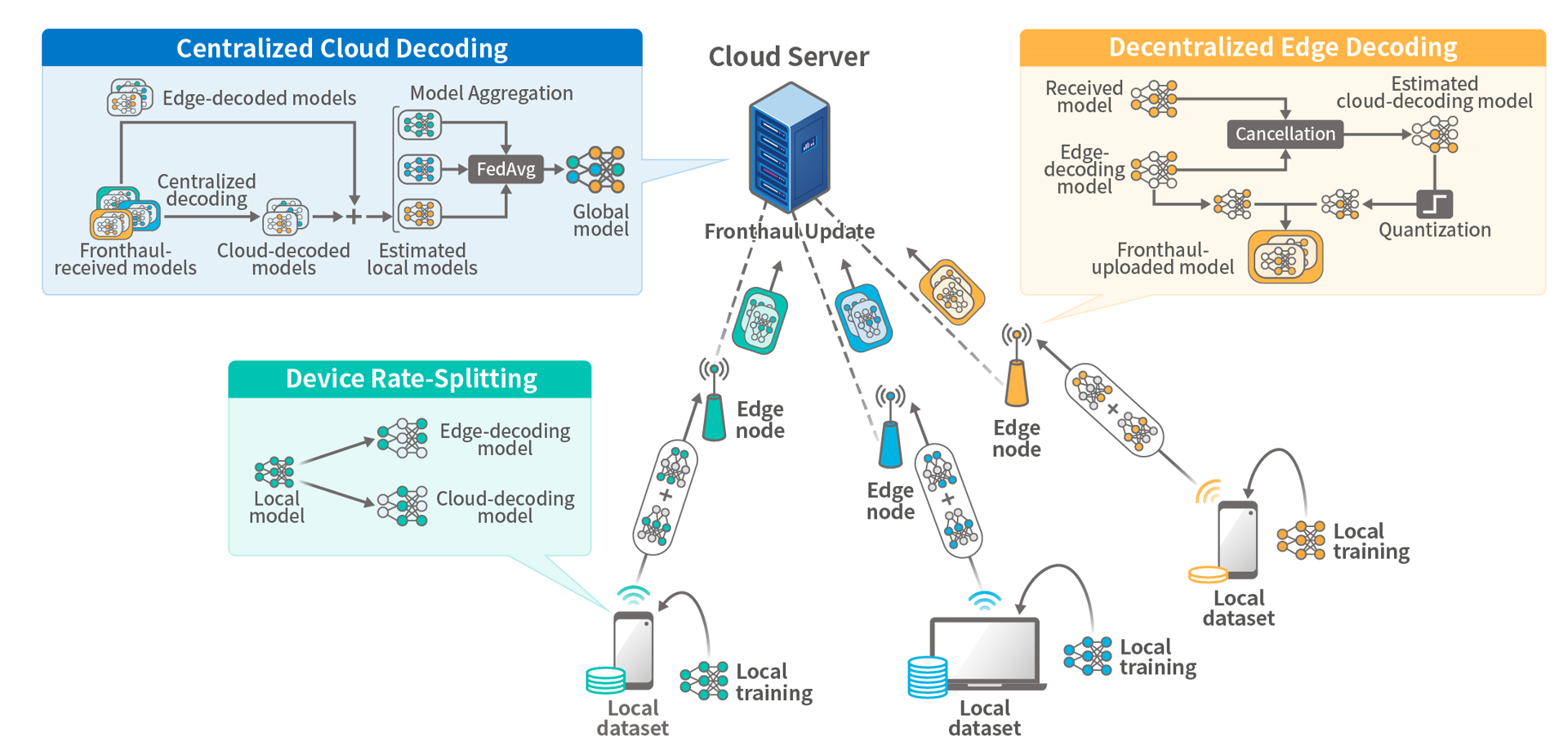} \vspace{-0.1cm}\caption{Rate-splitting-based wireless federated learning system.} \label{fig:FL} \vspace{-0.3cm}
\end{figure*}

Wireless federated learning (FL) has received regarded as a key enabler of remote machine learning systems for distributed wireless devices of 6G. A promising network architecture for the FL is fog radio access networks (F-RANs) which leverage proximity edge nodes (ENs) to collect local model parameters of wireless devices \cite{zhao:wirelessmag:20}. For the global model aggregation at the cloud, fronthaul interaction is essential to forward local models received by the ENs. Finite-capacity fronthaul channels invoke fundamental challenges in designing F-RAN-assisted FL protocols. There are two different strategies according to the decoding destinations: cloud-decoding and edge-decoding schemes \cite{park:tvt:22}. In the cloud-decoding approach, the cloud estimates local models with the aid of centralized signal processing algorithms. To this end, ENs upload quantized received signals via capacity-constrained fronthaul links, thereby inducing quantization errors in the model aggregation step. On the contrary, the edge-decoding scheme allows ENs to recover local models of associated devices. The resultant models are represented by digital values, e.g., float32-type tensors, which can be readily transferred through finite-capacity fronthaul links. However, individual edge-decoding processes incur severe inter-device interference and degrade the model accuracy.

RSMA can be employed to integrate these two extreme approaches by leveraging its flexibility in message construction \cite{park:tvt:22}. As illustrated in Fig. \ref{fig:FL}, each device first splits its message into edge-decoding and cloud-decoding parts. ENs individually recover their messages to cancel the model interference in the received signals. The resultant signal is quantized to accommodate the finite-capacity fronthaul links. The cloud recovers the cloud-decoding messages and integrates them with the edge-decoded models to reconstruct all the local model parameters. A key enabler of this approach is a joint optimization of precoding matrices of two split messages. This softly combines the edge-decoding and cloud-decoding approaches by adjusting transmit power levels between them. As a consequence, the rate-splitting-assisted FL system can achieve a better tradeoff between fronthaul overhead (communication cost) and learning accuracy.
{{Future work includes identifying a relationship between the message construction of RSMA and the global accuracy level in actual learning tasks. 
}}

\subsection{Non-Orthogonal Unicast and Multicast Transmissions}
With the explosive growth of multimedia applications over the mobile internet, such as live video streaming or gaming, the integration of multicast or broadcast  transmissions into existing unicast-only cellular networks has been investigated. 
To this end, 3GPP has standardized 5G Multimedia Broadcast Multicast Services (5G-MBMS) in Rel. 16. 
Aiming to ensure  efficient use of the available wireless resources by delivering a multicast service on top of a unicast network, non-orthogonal unicast and multicast (NOUM) transmission using the same spectrum and infrastructure has received increasing attention recently. 

RSMA inherently pushes multi-user transmission away from the conventional unicast-only transmission to non-orthogonal unicast and multicast transmission for the purposes of interference management with the help of SIC  at the receiver. 
Exploiting the flexibility of RSMA, RSMA-aided NOUM  has been proposed \cite{mao:tcom:19}. To be more specific, the unicast message is split into common and private parts at the transmitter, and the common parts are encoded with a multicast message into a super-common message for all users. 
By doing so, the SIC operation should be designed not only for managing the interference between unicast transmissions due to the imperfect CSIT but also managing the interference between multicast and unicast transmissions via NOUM.
{{
While most of the existing works are limited to the simplest case where a multicast message is intended for all users, i.e., a single multicast group, a joint design for multi-group multicast and unicast transmission is worth investigating in  future works.}}


\section{Summary and Outlook}


In the sense that multi-user systems are natural to modern wireless networks, MA is very critical to achieve the high performance required in 6G. Fundamentally, MA is related to interference management, and this viewpoint leads us to explore RSMA which embraces the conventional types of interference management strategies. 
To shed light on RSMA as a promising 6G MA technique, we have first explained the theoretical foundation of RSMA, including the key distinguishable features from conventional NOMA and multicasting, and the information-theoretical rationale for RSMA performance gains. Subsequently, we have elucidated four specific benefits that RSMA can provide depending on various scenarios, and how each benefit of RSMA enables ten promising 6G applications and scenarios. 
Specifically, the RSMA benefits and the corresponding 6G applications and scenarios are: 1) {\it{spectral efficiency}} (Grant-free massive mMTC, ISAC, and low-resolution quantizers), 2) {\it{robustness}} (mmWave \& THz communications, IRS, and NTN), 3) {\it{scalability}} (CFmMIMO, C-RAN), and 4) {\it{flexibility}} (wireless FL and NOUM). Through these, we assert that RSMA is indispensable MA for paving the way towards 6G networks. 

\bibliographystyle{IEEEtran}
\bibliography{reference.bib}

\end{document}